\documentclass{pos}
\usepackage{amsmath}
\usepackage{amssymb}%
\usepackage{amsfonts}%
\usepackage{graphicx}

\title{Thermal ground state for pure SU(2) Yang-Mills thermodynamics}
\ShortTitle{A thermal ground state for SU(2) Yang-Mills theory and the trace anomaly}
\author{\speaker{Francesco Giacosa}\\
Institute for Theoretical Physics, Goethe University,\\
Max-von-Laue-Str. 1, Frankfurt am Main, Germany\\
e-mail: giacosa@th.physik.uni-frankfurt.de}

\abstract{In this proceeding the emergence of a composite, adjoint-scalar field as an average over 
(trivial holonomy) calorons and anti-calorons is reviewed. This composite field acts 
as a background field to the dynamics of perturbative gluons, to which it is coupled via an
effective, gauge invariant Lagrangian valid for temperatures above the 
deconfinement phase transition. Moreover a Higgs mechanism is induced by the composite field: two gluons 
acquire a quasi-particle thermal mass. On the phenomenological side the composite field acts as a 
bag pressure which shows a linear dependence on the temperature. As a result the linear rise 
with temperature of the trace anomaly is obtained and is compared to recent lattice studies. }

\FullConference{8th Conference Quark Confinement and the Hadron Spectrum \\
                 September 1-6 2008\\
                 Mainz, Germany}

\begin{document}

\section{Introduction}

The SU(2) Yang-Mills (YM) theory at nonzero temperature is subject both of
theoretical and numerical on-going efforts. The aim is a deeper understanding
of its non-perturbative properties, which makes this theory complex and rich.
This is also a necessary step toward the understanding of the quark gluon
plasma (QGP), see for instance \cite{dirk} for a review. In this work, based
on Refs. \cite{ralfrev,herbst,garfield,lingrow,landau}, the
non-perturbative sector of SU(2) is described by a composite, (adjoint-)scalar
field $\phi$ in the deconfined phase ($T>T_{c})$.

Topological objects named calorons (i.e. instantons at nonzero $T$) with zero
and nonzero holonomy -the latter being a property of the fields at spatial
infinity- have been described in \cite{hist}. A non-trivial holonomy caloron
carries also monopole-antimonopole constituents. The `composite' field $\phi,$
which we want to introduce, emerges as an `average' over calorons and
anticalorons with trivial holonomy, see \cite{ralfrev,herbst} for a microscopic
derivation and \cite{garfield} for a macroscopic one. It depends only on the
temperature $T$ and the YM-scale $\Lambda$. On a length scale
$l>\left\vert \phi\right\vert ^{-1}$ it is thermodynamically exhaustive to
consider only the average field $\phi$ and neglect the (unsolvable)
microscopic dynamics of all YM-field configurations, such as calorons and
monopoles. One can then build up an effective theory for YM-thermodynamics
valid for $T>T_{c}$, in which the scalar field $\phi$ acts as background field
coupled to the residual, perturbative gluons. It also acts as an Higgs-field
in the thermal medium, implying that two gluons (out of three) acquire a
non-vanishing quasi-particle thermal mass. On a phenomenological level it
contributes to the energy and pressure as a temperature-dependent bag
constant. Here we shall focus on one particular implication of this effective
description: the linear growth with $T$ of the stress-energy tensor, which has
been obtained in some analyses of lattice data.

\section{Linear growth of $\theta=\rho-3p$ and bag constant}

Be $\rho$ the energy density and $p$ the pressure of a system at a given
temperature $T$. The quantity $\theta=\rho-3p$ is the trace of the
stress-energy tensor and vanishes for a conformal theory (as, for instance, a
gas of photons). In SU(2) this symmetry is broken by quantum effects (trace anomaly). In
\cite{miller}, based on the new lattice data of Ref. \cite{boyd}, it is found that
$\theta$ growths linearly with $T$:
\begin{equation}
\theta=aT,%
\begin{array}
[c]{cc}
&
\end{array}
2T_{c}\lesssim T\lesssim5Tc,%
\begin{array}
[c]{c}
\end{array}
\text{ }a\simeq1.5\text{ GeV}^{3}.
\end{equation}
(We notice that in the analysis of the same lattice data a quadratic rise of
$\theta,$ rather than a linear one, has been found \cite{pisarski}.) We also
refer to \cite{Kallman:1984ky}, where a simple linear fit $\theta=aT$ was found to
reproduce old lattice results. Recently, in \cite{bugaev} the linear rise has
been confirmed by studying the Lattice data of \cite{karsch}.

We now turn to a phenomenological description of a plasma of quasi-particles
\cite{levai}, where also a $T$-dependent bag is introduced to mimic a
non-perturbative behavior. As an example let us consider only one scalar field
with a $T$-dependent mass $m=m(T)$ and a bag $B=B(T)$. The energy density and
the pressure read (see, for instance, \cite{weise}):%
\begin{equation}
\rho=\rho_{p}+B(T),%
\begin{array}
[c]{c}
\end{array}
\text{ }p=p_{p}-B(T),
\end{equation}%
\[
\rho_{p}=\int_{k}\frac{\sqrt{k^{2}+m^{2}(T)}}{\exp\left[  \frac{\sqrt
{k^{2}+m^{2}(T)}}{T}\right]  -1},%
\begin{array}
[c]{c}
\end{array}
\text{ }p_{p}=-T\int_{k}\log\left[  1-\exp\left[  -\frac{\sqrt{k^{2}+m^{2}%
(T)}}{T}\right]  \right]
\]
where $\int_{k}=\int\frac{d^{3}k}{(2\pi)^{3}}.$ Requiring the validity of the
thermodynamical self-consistency \cite{weise,rischke} $\rho=T(dp/dT)-p$ (which
is a consequence of the first principle of thermodynamics), we obtain the
equation
\begin{equation}
\frac{dB}{dT}=-D(m)\frac{dB}{dT},%
\begin{array}
[c]{c}
\end{array}
\text{ }D(m)=\int_{k}\frac{m}{\sqrt{k^{2}+m^{2}}}\frac{1}{\exp\left[
\frac{\sqrt{k^{2}+m^{2}(T)}}{T}\right]  -1}.
\end{equation}
Imposing that $B(T)=cT$ and following the analytical steps of Ref.
\cite{lingrow}, we obtain:
\begin{equation}
\theta=\rho-3p=4B+\rho_{p}-3p_{p}\overset{T\text{ large}}{=}6B(T)=6cT.
\end{equation}
Thus, also the quasi-particle excitation contributes \emph{linearly} to
$\theta$ at high $T.$ The important result is that a linear rise of the bag
constant implies also a linear rise of $\theta=\rho-3p$. If the linear rise
shall be confirmed on the lattice, it means that the SU(2), nonperturbative
bag shall be a linear rising function with $T.$

\section{A thermal ground state}

The SU(2) (euclidean) YM Lagrangian reads $\mathcal{L}_{YM}=\frac{1}{2}Tr[G_{\mu\nu
}G^{\mu\nu}],$ where $G_{\mu\nu}=\partial_{\mu}A_{\nu}-\partial_{\nu}A_{\mu
}-ig[A_{\mu},A_{\nu}]$ and $A_{\mu}=A_{\mu}^{a}t^{a}$ ($t^{a}$ are the SU(2)
generators). $g$ is the fundamental coupling constant, which, upon
renormalization, is function of the renormalization scale $\mu.$ The
non-abelian nature of $\mathcal{L}_{YM}$ is at the origin of its
nonperturbative properties. A general field configuration can be decomposed as
$A_{\mu}=A_{\mu}^{\text{top}}+a_{\mu},$ where $A_{\mu}^{\text{top}}$ refers to
a topologically non-trivial function and $a_{\mu}$ to the quantum
fluctuations. It is very hard, if not impossible, to take into account at a
microscopic level all the topological objects such as calorons and monopoles.
As described in\ Refs. \cite{ralfrev,garfield} we introduce an adjoint scalar gauge field
$\phi=\phi^{a}t^{a}$ as%
\begin{equation}
\phi=\text{`Spatial average over (anti-)calorons with trivial holonomy'.}%
\end{equation}
The field $\phi$ can be univocally determined by the three following
conditions \cite{garfield}: (i) Being a spatial average it depends
(periodically) on $\tau$ only. It transforms as $\phi\rightarrow U\phi
U^{\dagger}$ under (space-independent) gauge transformations. (ii) The
corresponding Lagrangian $\phi$ reads $\mathcal{L}_{\phi}=Tr[(\partial_{\tau
}\phi)^{2}+V];$ BPS saturation of (anti-)calorons implies also BPS-saturation
for $\phi$: $\mathcal{H}_{\phi}=Tr[(\partial_{\tau}\phi)^{2}-V]=0.$ (iii)
$\phi$ represents (part of) the vacuum of the YM-system at nonzero $T$ and a
background field to the dynamics of the trivial quantum fluctuations $a_{\mu
}.$ The gauge-invariant quantity $\left\vert \phi\right\vert $ acts as a
`condensate' (strictly related to the gluon condensate, see \cite{landau}) and
does not correspond to any new particle. As a consequence $\left\vert
\phi\right\vert $ shall \emph{not} depend on $\tau.$

The conditions (i), (ii) and (iii) imply that $V(\phi)\varpropto1/\left\vert
\phi\right\vert ^{2}$ \cite{garfield}. Introducing a scale $\Lambda,$ which is
the only free parameter of the theory and is naturally identified with the
YM-scale, we obtain $V(\phi)=\frac{\Lambda^{6}}{\left\vert \phi\right\vert
^{2}}$. Solving the equation of motion one obtains as a unique solution (up to
a phase) $\left\vert \phi\right\vert =\sqrt{\Lambda^{3}/(2\pi T)}.$ On the
length scale $l>\left\vert \phi\right\vert ^{-1}$ (part of) the field
configurations $A_{\mu}^{\text{top}}$ are effectively described by $\phi.$

As a last step we couple $\phi,$ which is a given background solution
$\left\vert \phi\right\vert =\sqrt{\Lambda^{3}/(2\pi T)}$, to the quantum
fluctuations $a_{\mu}$ in a gauge invariant manner and obtain the final form
of the effective theory:
\begin{equation}
\mathcal{L}^{\text{eff}}\mathcal{=}Tr\left[  F_{\mu\nu}F^{\mu}+(D_{\mu}%
\phi)^{2}+\frac{\Lambda^{6}}{\left\vert \phi\right\vert ^{2}}\right]
,\label{effth}%
\end{equation}
with $F_{\mu\nu}=\partial_{\mu}a_{\nu}-\partial_{\nu}a_{\mu}-ie[a_{\mu}%
,a_{\nu}]$ and $D_{\mu}\phi=\partial_{\mu}\phi-ie[A_{\mu},\phi].$ The new,
$T$-dependent coupling constant $e=e(T)$ is univocally obtained by imposing
thermodynamical self-consistency as described in Section 2. In Ref. \cite{landau}
the fundamental coupling constant $g$ is related to $e(T)$ and it is found
that also $g$ admits a Landau pole, which is just slightly shifted from the
perturbative result. The effective coupling constant $e(\lambda=\frac{2\pi
T}{\Lambda})$ together with some relevant thermodynamical quantities evaluated
at tree-level in Fig. 1. Note that $e(\lambda)$ diverges at $\lambda
_{c}=13.86,$ corresponding to a critical temperature of $T_{c}\sim2\Lambda.$
Thus the effective theory is valid for $T>T_{c}.$ For $T>>T_{c}$ a plateaux is
reached for $e(\lambda)\sim\sqrt{8}\pi.$ Moreover, as evident from eq.
(\ref{effth}) a Higgs mechanism takes place: two gluons acquire a
quasi-particle mass $m(T)=2e(T)\left\vert \phi\right\vert $ which diverges at
the phase boundary.%
\begin{figure}
[ptb]
\begin{center}
\includegraphics[
height=1.7417in,
width=5.1275in
]%
{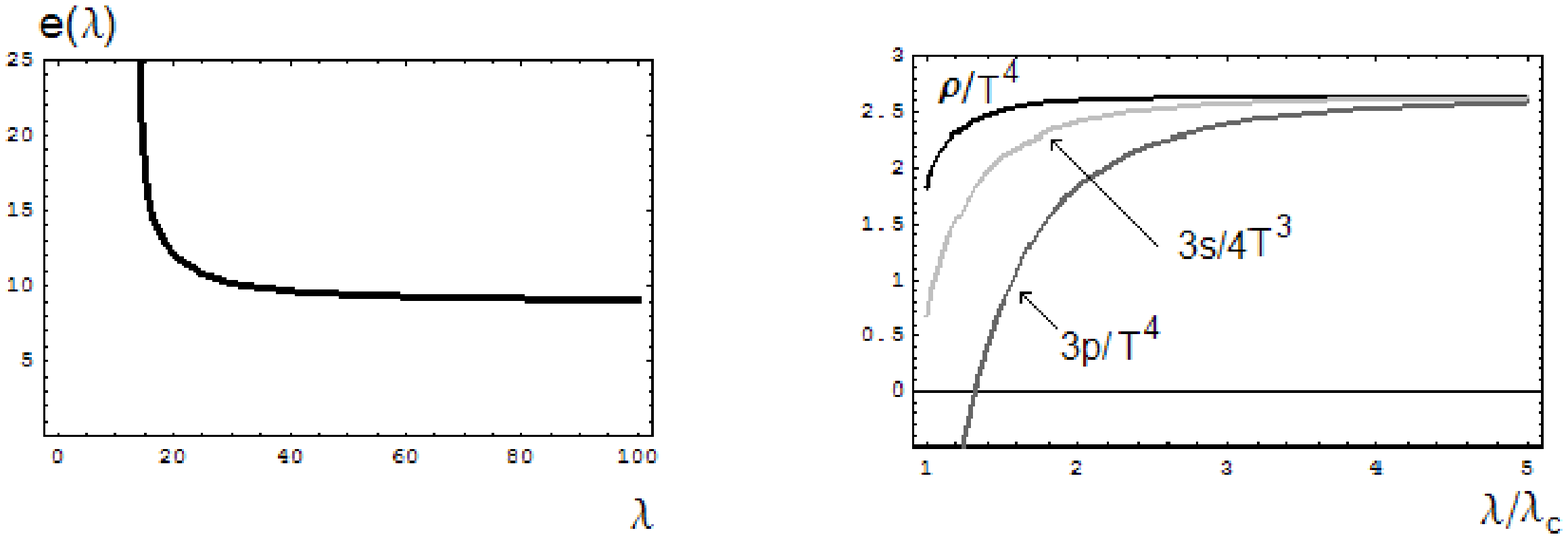}%
\caption{Left: effective coupling as function of $\lambda=2\pi T/\Lambda.$
Right: (scaled) thermodynamical relevant quantities as function of
$\lambda/\lambda_{c}=T/T_{c}.$}%
\end{center}
\end{figure}

Finally, we summarize the effective theory in Fig. 2. Its tree-level equations
are similar to those presented in Section 2. Corrections beyond tree-level are
small (1\%) \cite{markus}.%

\begin{figure}
[ptb]
\begin{center}
\includegraphics[
height=2.5763in,
width=3.9885in
]%
{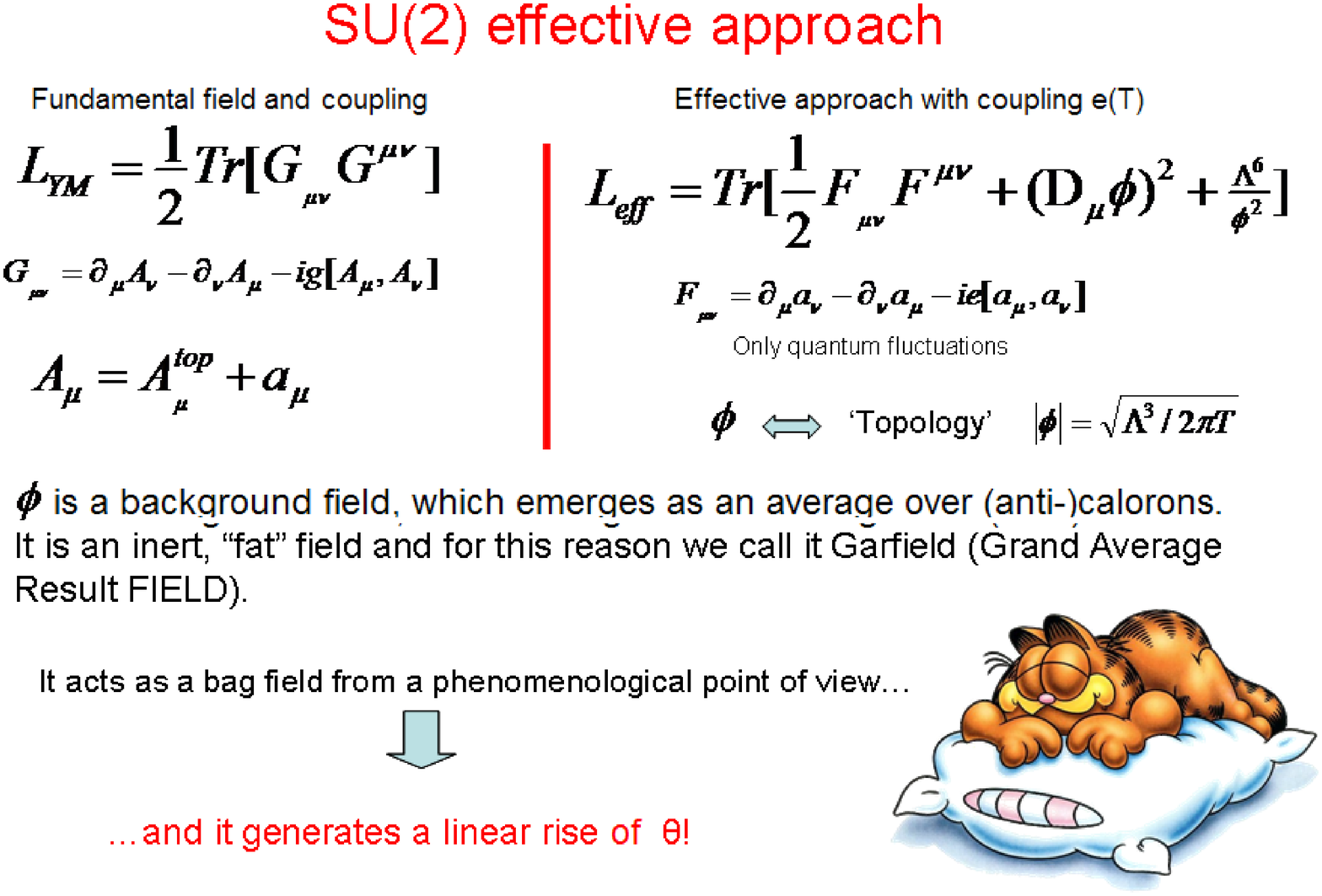}%
\caption{Summary of the effective theory: a transparency of the talk at
Confinement08 in Mainz.}%
\end{center}
\end{figure}

\section{Result for $\theta$ and conclusions}

The last term of eq. (\ref{effth}) acts as a $T$-dependent bag term,
$B(T)\sim\frac{\Lambda^{6}}{\left\vert \phi\right\vert ^{2}}=2\pi T\Lambda
^{3}.$ The important point is that $B(T)$ is linear in $T!$ It then implies,
as seen in Section 2, that $\theta$ of SU(2) YM-theory grows linearly in $T$
for high $T$. The precise result can be obtained analytically \cite{lingrow}:%
\begin{equation}
\theta=\rho-3p\overset{T>2T_{c}}{\sim}24\pi\Lambda^{3}T\simeq(1.7\text{
GeV}^{3})T.
\end{equation}
Note that the coefficient $1.7$ GeV$^{3}$ is similar to $1.5$ GeV$^{3}$ found
in Ref. \cite{miller} and is also not far from the theoretical result of Ref.
\cite{zwanziger}. We also note that the same approach can be applied to the
$SU(3)$ case, where one also finds $\theta\overset{T>2T_{c}}{\sim}24\pi
\Lambda^{3}T.$

We conclude this brief report on an effective approach for the description of
YM at nonzero $T$ by summarizing its main idea: the introduction of a simple,
average background field over calorons and anticalorons. The aim is a to avoid
an impossible, microscopic evaluation of complicated field configurations, and
to obtain a thermodynamically exhaustive description of a YM system. A simple
consequence of this effective approach, namely the linear growth with $T$ of
the trace anomaly, is in agreement with recent lattice simulations. Some
applications of this approach can be found in Refs. \cite{extra}.

\acknowledgments
The author thanks R. Hofmann and M. Schwarz for a fruitful collaboration.

\end{document}